\documentclass[english,aps,twocolumn,showpacs,superscriptaddress]{revtex4}
\usepackage[T1]{fontenc}
\usepackage[latin9]{inputenc}
\usepackage{xcolor}
\usepackage{pdfcolmk}
\usepackage{babel}
\usepackage{amsmath}
\usepackage{amssymb}
\usepackage{mathdots}
\usepackage{graphicx}
\usepackage{esint}
\PassOptionsToPackage{normalem}{ulem}
\usepackage{ulem}
\PassOptionsToPackage{version=3}{mhchem}
\usepackage{mhchem}
\usepackage[unicode=true,pdfusetitle,
 bookmarks=true,bookmarksnumbered=false,bookmarksopen=false,
 breaklinks=false,pdfborder={0 0 1},backref=false,colorlinks=false]
 {hyperref}

\makeatletter

\providecolor{lyxadded}{rgb}{0,0,1}
\providecolor{lyxdeleted}{rgb}{1,0,0}

\@ifundefined{textcolor}{}
{%
 \definecolor{BLACK}{gray}{0}
 \definecolor{WHITE}{gray}{1}
 \definecolor{RED}{rgb}{1,0,0}
 \definecolor{GREEN}{rgb}{0,1,0}
 \definecolor{BLUE}{rgb}{0,0,1}
 \definecolor{CYAN}{cmyk}{1,0,0,0}
 \definecolor{MAGENTA}{cmyk}{0,1,0,0}
 \definecolor{YELLOW}{cmyk}{0,0,1,0}
}

\makeatother

\begin{document}

\title{Threshold for Non-Thermal Stabilization of Open Quantum Systems}

\author{C. Y. Cai}

\affiliation{State Key Laboratory of Theoretical Physics, Institute of Theoretical
Physics and University of the Chinese Academy of Sciences, Beijing
100190, People's Republic of China}

\author{Li-Ping Yang}

\affiliation{State Key Laboratory of Theoretical Physics, Institute of Theoretical
Physics and University of the Chinese Academy of Sciences, Beijing
100190, People's Republic of China}

\author{C. P. Sun}

\email{suncp@itp.ac.cn}

\homepage{http://www.csrc.ac.cn/ suncp/}

\affiliation{State Key Laboratory of Theoretical Physics, Institute of Theoretical
Physics and University of the Chinese Academy of Sciences, Beijing
100190, People's Republic of China}

\affiliation{Beijing Computational Science Research Center, Beijing 100084, China}
\begin{abstract}
We generally study whether or not the information of an open quantum
system could be totally erased by its surrounding environment in the
long time. For a harmonic oscillator coupled to a bath of a spectral
density with zero-value regions, we quantitatively present a threshold
of system-bath coupling $\eta_{c}$ , above which the initial information
of the system can remains partially as its long time stablization
deviates from the usual thermalization. This non-thermal stabilization
happens as a non-Markovian effect. 
\end{abstract}

\pacs{03.65.Yz, 03.67.-a, 05.70.Ln, 42.50.Lc}

\maketitle
\emph{Introduction.---}Thermalization is a dynamic process of an open
system reaching the thermal equilibrium at the same temperature $T$
as its surrounding heat bath. From the point of view of the information
theory, thermalization is regarded as an information erasure process~\cite{MD_book}.
The open system initially prepared in an arbitrary state will relax
to a thermal state after a long-time Markov process. This steady state
is irrelevant to the system's initial state at all, but it carries
partial bath's information characterized its temperature $T$. Thus,
the conventional thermalization plays a necessary role in the initializations
of computation or thermodynamic cycle~\cite{Dong2011,Plenio2001,Quan2006}.
This perspective results in a comprehensive understanding for Landauer's
erasure principle~\cite{Landauer1961,MD_book}.

Thermalization is dynamically associated with the Markovian processs~\cite{Breuer2002}
and also can be described by the Langevin equation under the Wigner-Weisskopf
approximation~\cite{Leggett1983,YuLH1994}. However, it was found
that a strong system-bath coupling might result in a non-Markovian
process when the interection spectral density has zero-value regions~\cite{XiongHN2010,YangLP2012,ZhangWM2012,Breuer2009,Znidaric2011,ZhangWM_Opt2011}.
Two questions naturally follow for further investigation: (1) to what
extent the strength of system-bath coupling increases so that the
system's stabilization largely deviates from the usual thermalization?
(2) how much information of the initial state is left in the final
stable state for a non-Markovian process?

To answer these questions generally, we revisit the ``standard model''
of open quantum system, a harmonic oscillator (HO) coupled to a bath
of HOs with a spectral density with zero-value regions. We analytically
examine the mean occupation number of the system through the formally
exact solution to the Heisenberg equation of the total system. The
system's mean occupation number is divided into two parts, one of
which only depends on the system's initial state, and the other depends
on the environment at tememperature $T$. By a detailed asymptotic
analysis, we find that the first part does not vanish even for an
infinitely long time if the system-bath coupling strength exceeds
a threahold $\eta_{c}$ , which depends on the structure of the interaction
spectral density. Finally, we analyse the long-time behavior of the
second part of the mean occupation number, which depends on the population
distribution of the bath mode, to show how much information the system
inherits from the bath.

\emph{The ``standard model'' of open quantum system}.\emph{---}We
consider an open system consisting of a harmonic oscillator interacting
with its environment (or bath). The environment is modeled as a collection
of harmonic oscillators with linear coupling to the system. This has
been extensively studied in numerous literatures as a\emph{ }``standard
model'' of open quantum system, since it can be universally utilized
to reveal the core spirit of quantum dissipation process according
to Caldeira and Leggett~\cite{Leggett1983}.

The total Hamiltonian of our model reads 
\begin{equation}
H=\Omega\mathrm{a}^{\dagger}\mathrm{a}+\sum_{l}\omega_{l}\mathrm{b}_{l}^{\dagger}\mathrm{b}_{l}+\sum_{l}\left(\eta_{l}\mathrm{a}^{\dagger}\mathrm{b}_{l}+\eta_{l}^{*}\mathrm{b}_{l}^{\dagger}\mathrm{a}\right),
\end{equation}
where $\mathrm{a}(\mathrm{a}^{\dagger})$ and $\mathrm{b}_{l}(\mathrm{b}_{l}^{\dagger})$
are the annihilation(creation) operators of the system and the $l$-th
mode of the environment, respectively. The corresponding Heisenberg
equation has the formal solution~\cite{Louisell,Sun1998}, 
\begin{equation}
\mathrm{a}(t)=u(t)\mathrm{a}+\sum_{l}u_{l}(t)\mathrm{b}_{l}.
\end{equation}
The coefficient $u(t)$ is governed by the following differential-integral
equation, 
\begin{equation}
\frac{\mathrm{d}u(t)}{\mathrm{d}t}+\mathrm{i}\Omega u(t)+\int_{0}^{t}G(t-\tau)u(\tau)\mathrm{d}\tau=0,\label{eq:u_t}
\end{equation}
with the initial condition $u(0)=1$. Here, the integral kernel 
\begin{equation}
G(t)=\mathcal{F}[J(\omega)]\equiv\frac{1}{2\pi}\int_{-\infty}^{\infty}J(\omega)\mathrm{e}^{-\mathrm{i}\omega t}\mathrm{d}\omega,
\end{equation}
is the Fourier transformation of the system-bath interaction spectral
density $J(\omega)\equiv2\pi\sum_{l}\left|\eta_{l}\right|^{2}\delta(\omega-\omega_{l})$,
which is usually taken as \emph{a priori }microscopic knowledge. The
other coefficients $u_{l}(t)$ are given by 
\begin{equation}
u_{l}(t)=-\mathrm{i}\eta_{l}\int_{0}^{t}u(t-\tau)\mathrm{e}^{-\mathrm{i}\omega_{l}\tau}\mathrm{d}\tau.\label{eq:u_l}
\end{equation}

When the system and the bath are initially in the direct product state
$\rho(0)=\rho_{S}(0)\otimes\rho_{E}(0)$, where $\rho_{E}(0)$ is
the thermal equilibrium state of the bath at temperature $T$ and
$\rho_{S}(0)$ is an arbitrary initial state of the system, the system's
mean occupation number is obtained as~\cite{tan2011} 
\begin{equation}
n(t)=\left|u(t)\right|^{2}\left\langle \mathrm{a}^{\dagger}\mathrm{a}\right\rangle _{S}+\sum_{l}\left|u_{l}(t)\right|^{2}\left\langle \mathrm{b}_{l}^{\dagger}\mathrm{b}_{l}\right\rangle _{E},\label{eq:6}
\end{equation}
where $\langle\cdots\rangle_{S(E)}$ =${\rm Tr}_{S(E)}[\rho_{S(E)}\cdots]$
means the average over the state $\rho_{S(E)}$. The mean occupation
number $n(t)$ is devided into two parts: the first part, which vanishes
in a long time Markov process, only depends on the system's initial
condition. The second part, which usually leads to the thermalization
of the system in the weak-coupling case~\cite{Louisell}, charaterizes
the contribution from the thermal bath. In this letter, it will be
shown that the first part describes the dynamic process of erasing
or preserving the system's initial information, while the second part
describe how the bath's information is inherited by the system.

It is convenient to extend the limit of integration of $\tau$ in
Eq.~(\ref{eq:u_t}) from $[0,t]$ to $(-\infty,t]$ by defining $\left.u(t)\right|_{t<0}=0$,
and then the differential-integral equation (\ref{eq:u_t}) changes
into~\cite{method} 
\begin{equation}
\frac{\mathrm{d}u(t)}{\mathrm{d}t}+\mathrm{i}\Omega u(t)+\int_{-\infty}^{t}\mathrm{d}\tau G(t-\tau)u(\tau)=\delta(t).\label{eq:u_t-1}
\end{equation}
It is obvious that Eq.~(\ref{eq:u_t-1}) is exactly equivalent to
Eq.~(\ref{eq:u_t}) in the time domain $(0,\infty)$. A formal solution
of $u(t)$ is obtained via the Fourier transformation as 
\begin{equation}
u(t)=-\frac{1}{2\pi\mathrm{i}}\int\frac{\mathrm{e}^{-\mathrm{i}\omega t}\mathrm{d}\omega}{F(\omega)},\label{u}
\end{equation}
where the denominator in the integral is 
\begin{equation}
F(\omega)\equiv\omega-\Omega+\frac{1}{2\pi}\int\mathrm{P}\frac{J(\omega')\mathrm{d}\omega'}{\omega'-\omega}+\frac{\mathrm{i}}{2}J(\omega)+\mathrm{i}\epsilon,
\end{equation}
and $\epsilon$ is an infinitesimal positive constant. For some special
spectrum, e.g. a Lorentzian-type spectrum, the above integral can
be carried out analytically~\cite{ZhangWM2012}.

For $t\rightarrow\infty$, we assume an asymptotic solution of Eq.~(\ref{eq:u_t})
$u(t)\sim A\exp(-\mathrm{i}\omega_{0}t)$, which oscillates with a
single frequency $\omega_{0}$ and amplitude $A$. Due to the linearity
of Eq.~(\ref{eq:u_t}), the superposition of several such single-mode
solutions is also an asymptotic solution of Eq~(\ref{eq:u_t}). Therefore,
we only need to investigate the existence conditions and the properties
of the single-mode case. We first let $\tilde{u}(t)\equiv\exp(\mathrm{i}\omega_{0}t)u(t)$,
which satisfies a intergral-differential equation similar to Eq.(\ref{eq:u_t})
with modified frequency $\tilde{\Omega}\equiv\Omega-\omega_{0}$ and
modified kernel $\tilde{G}(t)\equiv\mathcal{F}[J(\omega+\omega_{0})]$~\cite{method}.
The steady asymptotic solution is determined by $\left.\mathrm{d}\tilde{u}(t)/\mathrm{d}t\right|_{t\rightarrow\infty}=0$,
or 
\begin{equation}
\!\!\!\!\!\left[\!\mathrm{i}(\Omega-\omega_{0})\!+\!\frac{1}{2\pi\mathrm{i}}\mathrm{P}\!\int_{-\infty}^{\infty}\!\!\frac{J(\omega\!+\!\omega_{0})}{\omega}\mathrm{d}\omega\!+\!\frac{1}{2}J(\omega_{0})\!\right]\!\!\cdot\!\! A\!=\!0.
\end{equation}
For the case with $A\not=0$, the above equation gives the criteria
for existence of nonvanishing solution of Eq.~(\ref{eq:u_t}) about
a real oscillating frequency $\omega_{0}$:\begin{subequations} 
\begin{eqnarray}
J(\omega_{0}) & = & 0,\label{criteria_1}\\
\Omega-\omega_{0} & = & \frac{1}{2\pi}\mathrm{P}\int_{-\infty}^{\infty}\frac{J(\omega)}{\omega-\omega_{0}}\mathrm{d}\omega.\label{criteria}
\end{eqnarray}
\end{subequations}

\emph{Criteria for non-thermal stabilization.---}In the conventional
thermalization process, $u(t)$ decays to $0$ as $t\rightarrow\infty$.
This effect implies that the system's initial information will be
totally erased. However, there exist some clues reminding us that
$u(t)$ may not vanishes at long time~\cite{XiongHN2010,YangLP2012,ZhangWM2012}.
Now, we explicitly present the criteria for the occurence of such
non-thermal stabilization.

According to Ref.~\cite{ZhangWM2012}, the non-thermal stabilization
firstly requires the spectrum $J(\omega)$ to have at least one zero-value
region. Thus, the non-thermal stabilization would never happen if
the spectrum were of Lorentzian-type. Eq.~(\ref{criteria}) must
have at least one solution in theses zero regions. If Eq.~(\ref{criteria})
has more than one solution in the zero-value regions of $J(\omega)$,
the general solution of Eq.~(\ref{eq:u_t}) will be the superposition
of these single-mode solutions.

\begin{figure}
\begin{centering}
\includegraphics[width=9cm]{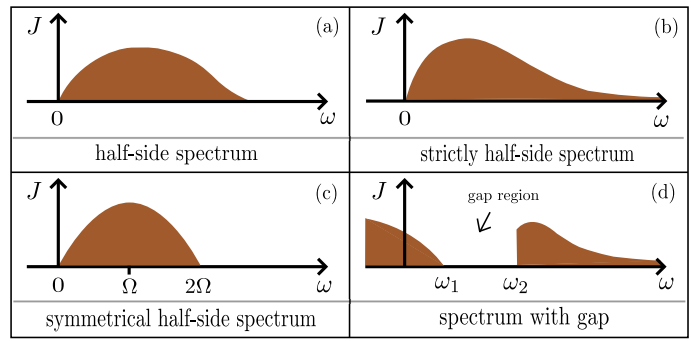} 
\par\end{centering}

\caption{\label{Spectrum}(color online). Four classes of spectrums studied
in this letter (a) half-side spectrum (b) strictly half-side spectrum
(c) symmetrical half-side spectrum (d) gapped spectrum}
\end{figure}

\begin{figure}
\begin{centering}
\includegraphics[width=9cm]{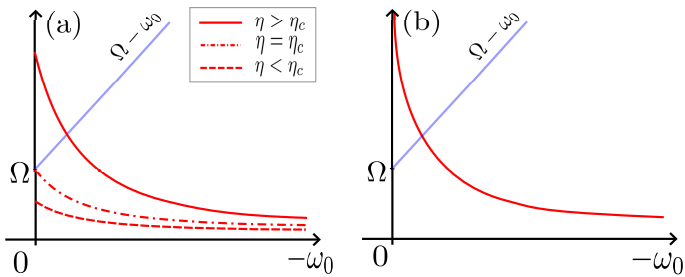} 
\par\end{centering}

\caption{\label{fig2}(color online). Schematic for the criterion (\ref{criteria})
for half-side spectrum. The blue lines represent the l.h.s. of Eq.(\ref{criteria})
as a function of $-\omega_{0}$. The red lines represent the r.h.s.
of Eq.(\ref{criteria}). The position where the red line crosses the
y-axis is mainly determine by $\eta$. (a) $\eta_{c}>0$. In this
case, there exists one solution for Eq.(\ref{criteria}) if $\eta\geqslant\eta_{c}$,
while there is no solution if $\eta<\eta_{c}$. (b) $\eta_{c}=0$.
In this case, there always exists a solution as long as the system
couples to the bath.}
\end{figure}

Let us consider a specific kind of spectrum that possesses a half-side
configuration$\left.J(\omega)\right|_{\omega<0}=0$ {[}see Fig.~\ref{Spectrum}(a){]}.
It should be emphasized that the interaction spectrum of a bosonic
bath is always of half-side form, otherwise the total Hamiltonian
will have no lower bound. Then, we consider whether there exists a
solution $\omega_{0}\ (\omega_{0}<0$) satisfying Eq.(\ref{criteria}).
The l.h.s. of Eq.(\ref{criteria}) is a monotonically increasing function
of $-\omega_{0}$ and has no upper limit, while the r.h.s. is a monotonically
decreasing function of $-\omega_{0}$ (see Fig.\ref{fig2}). Thus,
there is no more than one solution for Eqs.~(\ref{criteria_1}) and
(\ref{criteria}). Moreover, the crireria for the non-thermal stablization
reduces to 
\begin{equation}
\frac{1}{2\pi}\int_{0}^{\infty}\frac{J(\omega)}{\omega}\mathrm{d}\omega\geqslant\Omega.\label{criteria_halfside}
\end{equation}
Usually, the spectral density can be rewriten as $J(\omega)=\eta J_{0}(\omega)$,
where $\eta$ characterizes the system-bath interaction strength and
$J_{0}(\omega)$ describes the pure spectral structure. Thus, the
above condition (\ref{criteria_halfside}) becomes $\eta\geqslant\eta_{c}$,
where the threshold strength $\eta_{c}$ is 
\begin{equation}
\eta_{c}=2\pi\Omega\left(\int_{0}^{\infty}\frac{J_{0}(\omega)}{\omega}\mathrm{d}\omega\right)^{-1}.\label{eta_c}
\end{equation}

The above arguments show that , if the coupling strength $\eta<\eta_{c}$,
$u(t)$ asymptotically would vanish as $t\rightarrow\infty$. Previously,
this quantitative criteria by $\eta_{c}$ was qualitatively described
by sentance ``coupling is weak enough''. When the coupling strength
is so strong that $\eta\geqslant\eta_{c}$, the asymptotic value of
$\left|u(t)\right|\not=0$ and then the intitial information of the
system will not be totally erazed enven at long time. Consequently,
the Markov approximation can not work well when $\eta>\eta_{c}$.
When the half-side spectral density satisfies $\int_{0}^{\infty}[J_{0}(\omega)/\omega\mathrm{]d}\omega=\infty$,
the critical coupling strength becomes zero according to Eq.~(\ref{eta_c})
{[}see Fig.\ref{fig2}(b){]}. Thus, no matter how weak the system-bath
interaction is, the stabilization is non-thermal and the Markov approximation
or Wigner-Weisskopf approximation is tnot valid. In other words, such
spectrum is born to be non-Markovian, e.g., the square spectrum.

Next, we show how to estimate the amplitude $A$ from the formally-exact
solution of $u(t)$ in Eq.(\ref{u}). For a given solution $\omega_{0}$
of the Eq.~(\ref{criteria_1}) and (\ref{criteria}), $F(\omega_{0})$
vanishes. Therefore, the integral around $\omega_{0}$ contributes
most to the integration in Eq.(\ref{u}) and $F(\omega)$ can be approximately
replaced by $F'(\omega_{0})(\omega-\omega_{0})$. According to the
residue theorem, we have $u(t)\simeq\exp(-\mathrm{i}\omega_{0}t)/F'(\omega_{0})$.
Then, the amplitude $A$ is approximated as $1/F'(\omega_{0})$~\cite{method},
i.e., 
\begin{eqnarray}
A & \simeq & \left(1+\frac{1}{2\pi}\int\mathrm{P}\frac{J(\omega)\mathrm{d}\omega}{(\omega-\omega_{0})^{2}}\right)^{-1}.\label{A}
\end{eqnarray}

\emph{Example of non-thermal stabilizations.---}The first example
of the non-thermal stabilization is the case with a symmetrical half-side
spectrum that satisfies $J(\Omega-\omega)=J(\Omega+\omega)$ with
respect to the resonace point $\omega=\Omega$ and $J(\omega)$ does
not vanish if and only if $\omega\in(0,2\Omega)$ (see Fig.\ref{Spectrum}c).
There exists a critical coupling strength $\eta_{c}$ determined by
Eq.~(\ref{eta_c})~\cite{method}. In the non-thermal region $\eta\geqslant\eta_{c}$,
the asymptotic solution of $u(t)$ is the superposition of two single-mode
solutions. In the supplemental material, we examine two concrete examples,
the triangular spectrum and the rectangular spectrum. The coincidence
between the analytical calculations and numerical results implies
that the criterion (\ref{criteria}) and the estimation of the amplitude
Eq.(\ref{A}) work well.

\begin{figure}
\begin{centering}
\includegraphics[width=9cm]{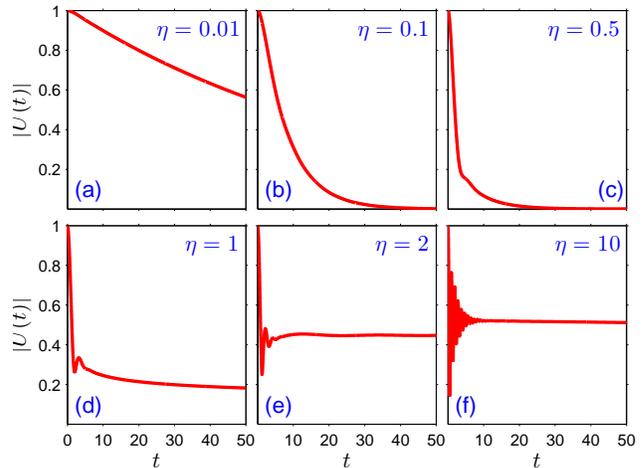} 
\par\end{centering}

\caption{\label{fig1}(color online). $\left|u(t)\right|$ as a function of
$t$ for Ohmic spectrum with coupling strength (1) $\eta=0.01$, (2)
$\eta=0.1$, (3) $\eta=0.5$, (4) $\eta=1$, (5) $\eta=2$, and (6)
$\eta=10$. The system's frequency is $\Omega=1$ and the spectrum
cutoff frequency is $\Omega_{c}=1$, which implies the critical coupling
strength $\eta_{c}=1$. }
\end{figure}

The second example is a more realistic one---Ohmic spectrum with the
density distribution: 
\begin{equation}
J(\omega)=2\pi\eta\theta(\omega)\omega\exp(-\omega/\Omega_{c}).
\end{equation}
Here, $\eta$ characterizes the coupling strength and $\Omega_{c}$
is the cutoff frequency. This spectrum is widely applied in open systems\cite{Leggett1987,Clerk}.
There exists a critial coupling strength $\eta_{c}=\Omega/\Omega_{c}$
according to Eq.(\ref{eta_c}). As shown by the numerical simulation
of $u(t)$ in Fig.\ref{fig1}(a-c), when $\eta<\eta_{c}$, $u(t)$
decays exponentially and the decay rate increases with $\eta$. On
the other hand, when $\eta\geqslant\eta_{c}$, $\left|u(t)\right|$
has a non-vanishing asymptotic value $|A|$, which increases with
the coupling strength $\eta$. The numerical results also confirms
the former qualitative analysis from Eq.(\ref{criteria}) and Eq.(\ref{A}).
Sometimes, the spectrum that one meets in the expriment may be a modified
one, such as sub-Ohmic or super-Ohmic spectrum. Our method can be
applied to these cases straightforwardly~\cite{ZhangWM2012}.

In the third example, the we deal with the gapped-spectrum case. As
shown in the supplementary, the density of the spectrum $J(\omega)$
vanishes if and only if $\omega\in(\omega_{1},\omega_{2})$ {[}see
Fig.\ref{Spectrum}(d){]}. One may meet this kind of spectrum when
the heat bath has a band gap, i.e., photonic crystal system\cite{photonic crystal,ZhangWM_Opt2011}.
One question arises that whether there exists a solution $\omega_{0}\in(\omega_{1},\omega_{2})$
satisfying the criterion (\ref{criteria}). After an argument similar
to the half-side spetrum case, we find that the criteria require the
frequency of the system $\Omega$ to satisfy $\Omega\in\left[\Omega_{1},\Omega_{2}\right]$,
where 
\begin{eqnarray*}
\Omega_{i} & \equiv & \omega_{i}-\frac{1}{2\pi}\int_{-\infty}^{\omega_{1}}\frac{J(\omega)}{\omega_{i}-\omega}\mathrm{d}\omega+\frac{1}{2\pi}\int_{\omega_{2}}^{\infty}\frac{J(\omega)}{\omega-\omega_{i}}\mathrm{d}\omega,
\end{eqnarray*}
for $i=1,2$. When the spectral density is discontinuous at $\omega_{1}$
and $\omega_{2}$, for example, 
\[
J(\omega)=\theta(\omega_{1}-\omega)\eta_{1}\mathrm{e}^{\gamma_{1}\omega}+\theta(\omega-\omega_{2})\eta_{2}\mathrm{e}^{-\gamma_{2}\omega},
\]
then $\Omega_{1}=-\infty$ and $\Omega_{2}=+\infty$. In this case,
the criterion $\Omega\in\left[\Omega_{1},\Omega_{2}\right]$ always
holds. Thus, no matter how weak the coupling strength is, such spectrum
is always associated with a non-vanishing asymptotic solution of $u(t)$
with oscillation frequency $\omega_{0}\in[\omega_{1},\omega_{2}]$.
Practically, it is useful to judge the existence and the location
of the spectral gap by measuring the frequency $\omega_{0}$.

\emph{The information} \emph{from the bath} \emph{inherited} \emph{by
system.---}We have described how the first part of the system's mean
occupation number can represent the residual information of the the
system's initial state. Now, we turn attention to the second part,
which depends on the population distribution of the bath.

The second part of the system's mean occupation number in Eq.~(\ref{eq:6})
is re-written as 
\begin{equation}
\sum_{l}\left|u_{l}(t)\right|^{2}\left\langle \mathrm{b}_{l}^{\dagger}\mathrm{b}_{l}\right\rangle =\int p(\omega)f_{\beta}(\omega)\mathrm{d}\omega,
\end{equation}
where $p(\omega)=\sum_{l}\left|u_{l}(t)\right|^{2}\delta(\omega-\omega_{l})$
and $f_{\beta}(\omega)=1/[\exp(\beta\omega)-1]$ with $\beta=1/(k_{B}T)$.
Actually, this part can be viewed as the information 'written' into
the system by the bath and $p(\omega)$ is the distribution function.
According to Eq.(\ref{eq:u_l}), $u_{l}(t)$ is determined by an integral
of $u(t)$ over the time domain $[0,t]$. As shown in the former sections,
$u(t)$ decays exponentially in short time and relaxes to an asymptotic
form $A\exp(-\mathrm{i}\omega_{0}t)$ at long time. In order to calculate
the second part of Eq.~(\ref{eq:6}), we consider two special cases:
(1) $A=0$ for small $\eta$, and (2) $A\not=0$ for large $\eta$.

The first case has been well studied~\cite{Louisell}. In this case,
$u_{l}(t)$ is dominanted by the short-time behaviour of $u(t)$ and
the distribution function $p(\omega)$ is approximated by a Lorentzian-type
distribution 
\begin{eqnarray}
p(\omega) & = & \frac{1}{2\pi}\cdot\frac{2\gamma}{(\omega-\Omega')^{2}+\gamma^{2}},
\end{eqnarray}
where the two parameters $\Omega'$ and $\gamma$ can be calculated
via the Wigner-Weisskopf approximation as shown in the supplemental
material. Because $A=0$, the first part of the system's mean occupation
number vanishes. In the weak coupling limit, $p(\omega)\rightarrow\delta(\omega-\Omega')$,
which leads to $n(T)\simeq f_{\beta}(\Omega').$ This implies that
the system's mean occupation number actually inherits the population
of the environment mode with the renormalized mode frequency $\Omega'$.

In the second case, $u_{l}(t)$ is dominated by the long-time behavior
of $u(t)$. If $u(t)$ has a single-mode asymptotic solution with
oscillating frequency $\omega_{0}$ when $t\rightarrow\infty$, the
distribution function is approximated as 
\begin{equation}
p(\omega)\simeq\frac{A^{2}}{2\pi}\cdot\frac{2J(\omega)}{(\omega-\omega_{0})^{2}}.
\end{equation}
This distribution is totally different from that in the weak coupling
case as the two following ways: 1. it is no longer normalized to unity.
This is a natural result since the system's mean occupation number
now depends on both the bath and its own initial value. Second, it
is a widespread distribution instead of a sharp one, which implies
that the information written by the environment becomes more complicated.
However, it should be emphasized that when temperature is low enough
($f_{\beta}(\omega)\rightarrow0$), the second term in Eq.(\ref{eq:6})
will be small compared to the first term. Thus, in this situation,
one may physically observe the non-thermal stabilization effect by
measuring the system's mean occupation number~\cite{experiment1}.

\emph{Remarks and conclusion.---}We have studied a non-thermal stabilization
phenomenon by calculating the open system's mean occupation number.
The criteria for this non-Markvoian effect was presented with a quantitative
threshold $\eta_{c}$ for most system-bath interaction spectra. In
the non-thermal region, $\eta\geqslant\eta_{c}$, the system's initial
information of the system is no longer totally erased by the bath. 

Actually, $\eta_{c}$ explicitly provides the Markovian approximation
of quantum open system with a quantitative upper limit. Our investigation
undoublely clarified the misunderstanding that the Markovian approximation
is valid only when the coupling strength is small enough, which is
closely dependent of the structure of spectral density. In this sense
the non-thermal stabilization effect due to the non-Markovian proccess
above the threshold $\eta_{c}$ provides us with a new fashion to
understand the information lost in open systems. 

Apparentlly, our aproach is universal and can be applied to the Fermion
case. Then, the first open question is wether this non-thermal stabilization
could happen for a Fermion like system, such as two level atom coupling
to some bath. It is also worthy of discussing the impact of the non-thermal
stabilization on the entanglemen evolution~\cite{entangle_P,entangle_Z,entangle3}.

This work was supported by NSFC through grants 10974209 and 10935010
and by the National 973 program (Grant No. 2012CB922104).

\end{document}


\title{Supplementary information: Non-Thermal Stabilization of Open Quantum
Systems}

\author{C. Y. Cai}

\affiliation{State Key Laboratory of Theoretical Physics, Institute of Theoretical
Physics and University of the Chinese Academy of Sciences, Beijing
100190, People's Republic of China}

\author{Li-Ping Yang}

\affiliation{State Key Laboratory of Theoretical Physics, Institute of Theoretical
Physics and University of the Chinese Academy of Sciences, Beijing
100190, People's Republic of China}

\author{C. P. Sun}

\email{suncp@itp.ac.cn}

\homepage{http://www.csrc.ac.cn/~suncp/}

\affiliation{State Key Laboratory of Theoretical Physics, Institute of Theoretical
Physics and University of the Chinese Academy of Sciences, Beijing
100190, People's Republic of China}

\affiliation{Beijing Computational Science Research Center, Beijing 100084, China}

\maketitle

\section{Heisenberg equation and its formal solution}

We consider the {}``standard model'' of a quantum open system, a
harmonic oscillator with coordinate (momentum) $x\ (p)$, mass $M$,
and bare frequency $\tilde{\Omega}$, interacting with a bath of $N$
harmonic oscillators of coordinates $x_{l}$, mass $m_{l}$, and frequencies
$\omega_{l}$. The total Hamiltonian of the system and the bath is
\begin{equation}
H=\frac{p^{2}}{2M}+\frac{1}{2}M\tilde{\Omega}^{2}x^{2}+\sum_{l=1}^{N}\left(\frac{p_{l}^{2}}{2m_{l}}+\frac{1}{2}m_{l}\omega_{l}^{2}x_{l}\right)+\sum_{l=1}^{N}F_{l}(x)x_{l}+\Phi(x).
\end{equation}
Here, $\Phi(x)$ is added to concel the frequency renormalization.
In our case, the system-bath coupling is of a linear form, i.e., $F_{l}(x)=\tilde{\eta}_{l}x$.
The corresponding renormalization term is chosen as
\[
\Phi(x)=\sum_{l}\tilde{\eta}_{l}^{2}x^{2}/(2m_{l}\omega_{l}^{2}),
\]
to attain the minimum value of the potential energy of the total system
(system plus bath) for a given $x$. Now we obtained the renormalized
frequency (physical frequency) of the system
\begin{equation}
M\Omega^{2}=M\tilde{\Omega}^{2}+\sum_{l}\tilde{\eta}_{l}^{2}/(2m_{l}\omega_{l}^{2}).
\end{equation}

By taking the rotating wave approximation, we rewrite the model Hamiltonian
as 
\begin{equation}
H=\Omega\mathrm{a}^{\dagger}\mathrm{a}+\sum_{l}\omega_{l}\mathrm{b}_{l}^{\dagger}\mathrm{b}_{l}+\sum_{l}\left(\eta_{l}\mathrm{a}^{\dagger}\mathrm{b}_{l}+\eta_{l}^{*}\mathrm{b}_{l}^{\dagger}\mathrm{a}\right),
\end{equation}
where $\eta_{l}=\tilde{\eta}_{l}/\left(2\sqrt{Mm_{l}\Omega\omega_{l}}\right)$,
$a=\sqrt{M\Omega/2}[x+ip/(M\Omega)]$ and $\mathrm{b}_{l}=\sqrt{m_{l}\omega_{l}/2}[x_{l}+ip_{l}/(m_{l}\omega_{l})]$
are the annihilation operators of the system and the $l$-th mode
of the environment, respectively. The dynamics of the total system
is governed by the Heisenberg equation,
\begin{eqnarray}
\frac{\mathrm{d}\mathrm{a}(t)}{\mathrm{d}t} & = & -\mathrm{i}\Omega\mathrm{a}(t)-\mathrm{i}\sum_{l}\eta_{l}\mathrm{b}_{l}(t),\label{eq:2}\\
\frac{\mathrm{d}\mathrm{b}_{l}(t)}{\mathrm{d}t} & = & -\mathrm{i}\omega_{l}\mathrm{b}_{l}(t)-\mathrm{i}\eta_{l}^{*}\mathrm{a}(t).\label{eq:3}
\end{eqnarray}
The above equations are assumed to have the following formal solutions~\cite{Sun1998,Louiswell},
\begin{eqnarray}
\mathrm{a}(t) & = & u(t)\mathrm{a}+\sum_{l}u_{l}(t)\mathrm{b}_{l},\\
\mathrm{b}_{l}(t) & = & v_{l}(t)\mathrm{a}+\sum_{m}v_{lm}(t)\mathrm{b}_{m},
\end{eqnarray}
where $a$ and $b_{l}$ are the abbreviations for $a(0)$ and $b_{l}(0)$
respectively and the c-number time-dependent coefficients $u(t)$,
$u_{l}(t)$, $v_{l}(t)$, and $v_{lm}(t)$ are determined by equivalent
equations,
\begin{eqnarray}
\frac{\mathrm{d}u(t)}{\mathrm{d}t} & = & -\mathrm{i}\Omega u(t)-\mathrm{i}\sum_{l}\eta_{l}v_{l}(t),\label{eq:6}\\
\frac{\mathrm{d}u_{l}(t)}{\mathrm{d}t} & = & -\mathrm{i}\Omega u_{l}(t)-\mathrm{i}\sum_{m}\eta_{m}v_{ml}(t),\label{eq:7}\\
\frac{\mathrm{d}v_{l}(t)}{\mathrm{d}t} & = & -\mathrm{i}\omega_{l}v_{l}(t)-\mathrm{i}\eta_{l}^{*}u(t),\label{eq:8}\\
\frac{\mathrm{d}v_{lm}(t)}{\mathrm{d}t} & = & -\mathrm{i}\omega_{l}v_{lm}(t)-\mathrm{i}\eta_{l}^{*}u_{m}(t).\label{eq:9}
\end{eqnarray}
The initial conditions of Eq.~(\ref{eq:2}) and Eq.~(\ref{eq:3})
imply that $u(0)=1$, $u_{l}(0)=v_{l}(0)=0$, and $v_{lm}(0)=\delta_{lm}$.
The closed system formed by Eq.~(\ref{eq:6}) and Eq.~(\ref{eq:8})
straightforwardly gives~\cite{Louiswell}
\begin{equation}
\frac{\mathrm{d}u(t)}{\mathrm{d}t}=-\mathrm{i}\Omega u(t)-\int_{0}^{t}\sum_{l}\left|\eta_{l}\right|^{2}\mathrm{e}^{-\mathrm{i}\omega_{l}(t-\tau)}u(\tau)\mathrm{d}\tau.\label{eq:10}
\end{equation}
Usually, the total system is initially on a product state $\rho(0)=\rho_{S}(0)\otimes\rho_{E}(0)$,
where $\rho_{E}(0)$ is the thermal state of the bath at temperature
$T$. The mean occupation number of the system is obtained as~\cite{XiongHN2010}
\begin{equation}
n(t)=\left|u(t)\right|^{2}\left\langle \mathrm{a}^{\dagger}\mathrm{a}\right\rangle _{S}+\sum_{l}\left|u_{l}(t)\right|^{2}\left\langle \mathrm{b}_{l}^{\dagger}\mathrm{b}_{l}\right\rangle _{E},\label{eq:11}
\end{equation}
where $\langle\dots\rangle_{S(E)}={\rm tr}_{S(E)}[\rho_{S(E)}\dots]$. 

To study the information erasing process, we define the system-bath
interaction spectral density as
\begin{equation}
J(\omega)\equiv2\pi\sum_{l}\left|\eta_{l}\right|^{2}\delta(\omega-\omega_{l}),
\end{equation}
which is a non-negative function on the real $\omega$-axis and usually
is assumed to be \emph{a priori }microscopic knowledge. Then, the
dynamic equation (\ref{eq:10}) of $u(t)$ changes into~\cite{ZhangWM2012}
\begin{equation}
\frac{\mathrm{d}u(t)}{\mathrm{d}t}+\mathrm{i}\Omega u(t)+\int_{0}^{t}G(t-\tau)u(\tau)\mathrm{d}\tau=0,\label{du}
\end{equation}
where 
\begin{equation}
G(t)\equiv\frac{1}{2\pi}\int_{-\infty}^{\infty}J(\omega)\mathrm{e}^{-\mathrm{i}\omega t}\mathrm{d}\omega,
\end{equation}
is the time-domain representation of the spectral density $J(\omega)$.
Hereafter, $\int\mathrm{d}\omega$ denotes $\int_{-\infty}^{\infty}\mathrm{d}\omega$.
To extend the time domain from $[0,\infty)$ to $(-\infty,\infty)$,
we let $u(t)=\Theta(t)u(t)$, where $\Theta(t)$ is the step function.
It is found that $u(t)$ satisfies the following equation,
\begin{equation}
\frac{\mathrm{d}u(t)}{\mathrm{d}t}+\mathrm{i}\Omega u(t)+\int_{-\infty}^{t}\mathrm{d}\tau G(t-\tau)u(\tau)=\delta(t),
\end{equation}
with the new boundary condition $u(t)|_{t<0}=0$. Now it is very convenient
to use the Fourier-transform method to obtain the formal solution
of $u(t)$,
\begin{equation}
u(t)=-\frac{1}{2\pi\mathrm{i}}\int\frac{\mathrm{e}^{-\mathrm{i}\omega t}\mathrm{d}\omega}{\omega-\Omega+\frac{1}{2\pi}\int\mathrm{P}\frac{J(\omega')\mathrm{d}\omega'}{\omega'-\omega}+\frac{\mathrm{i}}{2}J(\omega)+\mathrm{i}\epsilon},\label{u}
\end{equation}
where $\epsilon$ is an infinitesimal positive constant.

\section{Criteria for non-thermal stabilization}

If the initial information of the system is not totally erased by
the bath when $t\rightarrow\infty$. We may assume that $u(t)$ is
asymptotically of the form 
\begin{equation}
u(t)\sim A\mathrm{e}^{-\mathrm{i}\omega_{0}t},
\end{equation}
where $A$ is the amplitude and $\omega_{0}$ is the single oscillating
frequency. Due to the linear homogeneity of Eq.~(\ref{eq:10}), the
superposition of several such single-mode-oscillation solution is
also its asymptotic solution. Consequently, we only need to study
the existence and the property of the single-mode one. To calculate
the oscillating frequency $\omega_{0}$, we let $\tilde{u}(t)\equiv\exp(\mathrm{i}\omega_{0}t)u(t)$
and get
\begin{equation}
\frac{\mathrm{d}\tilde{u}(t)}{\mathrm{d}t}+\mathrm{i}(\Omega-\omega_{0})\tilde{u}(t)+\int_{0}^{t}\mathrm{d}\tau\tilde{G}(t-\tau)\tilde{u}(\tau)=0,
\end{equation}
with the modified kernel 
\begin{equation}
\tilde{G}(t)\equiv\frac{1}{2\pi}\int_{-\infty}^{\infty}J(\omega+\omega_{0})\mathrm{e}^{-\mathrm{i}\omega t}\mathrm{d}\omega.
\end{equation}
The steady asymptotic solution is determined by $\left.\mathrm{d}\tilde{u}(t)/\mathrm{d}t\right|_{t\rightarrow\infty}=0$.
In this case, 
\begin{equation}
\lim_{t\rightarrow\infty}\mathrm{i}(\Omega-\omega_{0})\tilde{u}(t)=\mathrm{i}(\Omega-\omega_{0})A,
\end{equation}
and
\begin{eqnarray}
\lim_{t\rightarrow\infty}\int_{0}^{t}\mathrm{d}\tau\tilde{G}(t-\tau)\tilde{u}(\tau) & = & \lim_{t\rightarrow\infty}\int_{0}^{t}\mathrm{d}\tau\tilde{G}(\tau)\tilde{u}(t-\tau)\nonumber \\
 & \approx & \int_{0}^{\infty}\mathrm{d}\tau\tilde{G}(\tau)A\nonumber \\
 & = & \frac{1}{2\pi\mathrm{i}}\mathrm{P}\int_{-\infty}^{\infty}\frac{J(\omega+\omega_{0})}{\omega}\mathrm{d}\omega+\frac{1}{2}J(\omega_{0}).
\end{eqnarray}
Here, in the second step, we have used the fact that $\tilde{G}(\tau)$
decreases to zero rapidly when $\tau$ increases. Then, 
\begin{equation}
\left[\mathrm{i}(\Omega-\omega_{0})+\frac{1}{2\pi\mathrm{i}}\mathrm{P}\int_{-\infty}^{\infty}\frac{J(\omega+\omega_{0})}{\omega}\mathrm{d}\omega+\frac{1}{2}J(\omega_{0})\right]\cdot A=0.
\end{equation}
For the case $A\not=0$, we obtain the criteria for the existence
of the the nonvanishing solution of Eq.~(\ref{du}) with a real oscillating
freqency\begin{subequations}
\begin{eqnarray}
J(\omega_{0}) & = & 0,\\
\Omega-\omega_{0} & = & \frac{1}{2\pi}\mathrm{P}\int_{-\infty}^{\infty}\frac{J(\omega)}{\omega-\omega_{0}}\mathrm{d}\omega.\label{criteria}
\end{eqnarray}
\end{subequations}
\begin{figure}
\begin{centering}
\includegraphics[width=12cm]{spectrum1}
\par\end{centering}

\caption{\label{Spectrum}Four classes of spectrums studied in this paper (a)
half-side spectrum (b) Ohmic spectrum (c) symmetrical half-side spectrum
(d) gapped spectrum}
\end{figure}

For a special case, we now concern the half-side spectrum satisfying
$\left.J(\omega)\right|_{\omega<0}=0$ {[}see Fig.~\ref{Spectrum}(a){]}.
The l.h.s. of Eq.~(\ref{criteria}) is a monotonically increasing
function of $-\omega_{0}$ and have no upper limit, while the right-hand
side is a monotonically decreasing function of $-\omega_{0}$. The
real solution of Eq.~(\ref{criteria}) exists if and only if the
following condition is satisfied,
\begin{equation}
\frac{1}{2\pi}\int_{0}^{\infty}\frac{J(\omega)}{\omega}\mathrm{d}\omega\geqslant\Omega.\label{eq:26}
\end{equation}
Generally, we can use a single parameter $\eta$ to characterizes
the system-bath interaction strength with $J(\omega)=\eta J_{0}(\omega)$.
Then, the condition in Eq.~(\ref{eq:26}) becomes
\begin{equation}
\eta\geqslant\eta_{c},
\end{equation}
where the critical coupling strength $\eta_{c}$ is
\begin{equation}
\eta_{c}=2\pi\Omega\left(\int_{0}^{\infty}\frac{J_{0}(\omega)}{\omega}\mathrm{d}\omega\right)^{-1}.\label{eta_c}
\end{equation}

To estimate the amplitude $A$, we re-write the formally exact solution
$u(t)$ in Eq.~(\ref{u}) as 
\begin{equation}
u(t)=-\frac{1}{2\pi\mathrm{i}}\int\frac{1}{F(\omega)}e^{-\mathrm{i}\omega t}\mathrm{d}\omega.\label{eq:u1}
\end{equation}
Here,
\begin{equation}
F(\omega)\equiv\omega-\Omega+\frac{1}{2\pi}\int\mathrm{P}\frac{J(\omega')\mathrm{d}\omega'}{\omega'-\omega}+\frac{\mathrm{i}}{2}J(\omega)+\mathrm{i}\epsilon.
\end{equation}
For a given solution $\omega_{0}$ of Eq.~(\ref{criteria}), $F(\omega_{0})$
vanishes. Thus, the integral in the vicinity of $\omega_{0}$ in Eq.~(\ref{eq:u1})
is dominant and $F(\omega)$ can be approximatively replaced by its
first-order Taylor expansion $F(\omega)\approx F'(\omega_{0})(\omega-\omega_{0})$.
According to the residue theorem, we approximates $u(t)$ as
\begin{eqnarray}
u(t) & \simeq & \frac{1}{F'(\omega_{0})}e^{-\mathrm{i}\omega_{0}t}.
\end{eqnarray}
Then, the amplitude $A$ is approximated as $1/F'(\omega_{0})$, i.e.,
\begin{eqnarray}
A & \simeq & \left(1+\frac{1}{2\pi}\int\mathrm{P}\frac{J(\omega)\mathrm{d}\omega}{(\omega-\omega_{0})^{2}}\right)^{-1}.\label{A}
\end{eqnarray}

\section{Examples of non-thermal stabilizations}

\subsection{Symmetrical half-side spectrum}

As demonstrated in Fig.~\ref{Spectrum}(c), a symmetrical spectrum
$J(\omega)$ does not vanish if and only if $\omega\in(0,2\Omega)$
and $J(\Omega-\omega)=J(\Omega+\omega)$. There exists a critical
coupling strength $\eta_{c}$, which is determined by Eq.~(\ref{eta_c}).
When the coupling strength $\eta<\eta_{c}$, $u(t)$ vanishes as $t\rightarrow\infty$.
On the contrary, if $\eta\geqslant\eta_{c}$, $u(t)$ would have non-vanishing
asymptotic solutions. One of the solutions is $A\exp(-\mathrm{i}\omega_{0}t)$
with $\omega_{0}<0$. Due to the symmetry of the spectrum, one can
find another single-mode solution with the same amplitude $A$ and
the antipodal frequency $2\Omega-\omega_{0}$. Thus, the asymptotic
behavior of $u(t)$ is described by the superposition of the two single-mode
solutions,
\begin{equation}
U\sim2A\mathrm{e}^{-\mathrm{i}\Omega t}\cos(\Omega-\omega_{0})t,
\end{equation}
 where $\omega_{0}$ is determined by the criteria Eq.~(\ref{criteria})
and $A$ is estimated by Eq.~(\ref{A}).

We use these results to examine two kinds of spectrum as examples,
the triangle spectrum,
\begin{equation}
J_{1}(\omega)=\begin{cases}
2\pi\eta\frac{\omega}{\Omega} & 0\leqslant\omega\leqslant\Omega,\\
2\pi\eta\frac{2\Omega-\omega}{\Omega} & \Omega\leqslant\omega\leqslant2\Omega,\\
0 & \mathrm{otherwise},
\end{cases}
\end{equation}
and the square spectrum,
\begin{equation}
J_{2}(\omega)=\begin{cases}
2\pi\eta & 0<\omega<2\Omega,\\
0 & \mathrm{otherwise}.
\end{cases}
\end{equation}
According to Eq.~(\ref{eta_c}), the critical strength for the triangle
spectrum is $\eta_{c}=\Omega/(2\ln2)=0.7213\Omega$, while the critical
strength for the square spectrum is zero. The contrasts between analytical
calculations (starting from Eq.~(\ref{criteria}) and Eq.~(\ref{A}))
and numerical results (obtained by the numerical solution of Eq.~(\ref{du}))
are listed in Tab.~\ref{tab1} and Tab.~\ref{tab2}. By measuring
the oscillation period of $u_{0}(t)$, one finds out the oscillation
frequency $\Omega-\omega_{0}$. The coincidence between the analytical
calculations and numerical results implies that the criteria Eq.~(\ref{criteria})
and the estimation of the amplitude Eq.~(\ref{A}) indeed work well. 

\begin{table}
\begin{centering}
\begin{tabular}{|c|c|c|c|c|}
\hline 
$\eta$ & $\omega_{0}^{(a)}$ & $\omega_{0}^{(n)}$ & $2A^{(a)}$ & $2A^{(b)}$\tabularnewline
\hline 
\hline 
0.7321 & -0.0027 & -0.0026 & 0.4141 & $\sim$0.4\tabularnewline
\hline 
7.9577 & -1.8512 & -1.8501 & 0.9782 & $\sim$1.0\tabularnewline
\hline 
\end{tabular}
\par\end{centering}

\caption{\label{tab1}Contrast between analytical results $O^{(a)}$ and numerical
results $O^{(n)}$ for triangle spectrum with $\Omega=1$ and (1)
$\eta=0.7321$, (2) $\eta=7.9577$. }
\end{table}

\begin{table}
\begin{centering}
\begin{tabular}{|c|c|c|c|c|}
\hline 
$\eta$ & $\omega_{0}^{(a)}$ & $\omega_{0}^{(n)}$ & $2A^{(a)}$ & $2A^{(b)}$\tabularnewline
\hline 
\hline 
0.1 & -9.0722e-5 & -5.0721e-4 & 0.0018 & about 1e-3\tabularnewline
\hline 
0.5 & -0.1997 & -0.2006 & 0.6104 & about 0.6\tabularnewline
\hline 
50 & -9.0167 & -9.0264 & 0.9967 & about 1\tabularnewline
\hline 
\end{tabular}
\par\end{centering}

\caption{\label{tab2}Contrast between analytical results $O^{(a)}$ and numerical
results $O^{(n)}$ for square spectrum with $\Omega=1$ and (1) $\eta=0.1$,
(2) $\eta=0.5$, and (3) $\eta=50$. One should be noticed that the
actual observable in the numerical test is $\pi/(1-\omega_{0})$ instead
of $\omega_{0}$, thus $\omega_{0}^{(a)}$ is indeed very close to
$\omega_{0}^{(n)}$. }
\end{table}

\subsection{Ohmic spectrum}

For the Ohmic spectrum 
\begin{equation}
J(\omega)=2\pi\eta\theta(\omega)e^{-\omega/\Omega_{c}},
\end{equation}
there also exists a critical coupling strength $\eta_{c}=\Omega/\Omega_{c}$
according to Eq.~(\ref{eta_c}). When $\eta\geqslant\eta_{c}$, there
exists a single-mode asymptotic solution $A\exp(-\mathrm{i}\omega_{0}t)$
of $u(t)$ as $t\rightarrow\infty$. The oscillating frequency is
determined by the criteria (\ref{criteria}), 
\begin{equation}
\Omega-\omega_{0}=\eta\int_{0}^{\infty}\frac{\omega}{\omega-\omega_{0}}\mathrm{e}^{-\omega/\Omega_{c}}\mathrm{d}\omega,\label{eq:41}
\end{equation}
and the amplitude is
\begin{eqnarray}
A & \simeq & \left(\frac{\Omega-\eta\Omega_{c}}{\omega_{0}}-\frac{\Omega-\omega_{0}}{\Omega_{c}}\right)^{-1}.\label{eq:36}
\end{eqnarray}

\subsection{Gapped spectrum}

Now, we consider a gapped spectrum whose density vanishes if and only
if $\omega\in(\omega_{1},\omega_{2})$ as shown in Fig.~\ref{Spectrum}(d).
After a similar argument for deducing Eq.~(\ref{eq:26}), we discover
that if the eigen frequency of the system satisfies 
\begin{equation}
\Omega\in\left[\Omega_{1},\Omega_{2}\right],\label{eq:37}
\end{equation}
with 
\begin{eqnarray*}
\Omega_{1} & \equiv & \omega_{1}-\frac{1}{2\pi}\int_{-\infty}^{\omega_{1}}\frac{J(\omega)}{\omega_{1}-\omega}\mathrm{d}\omega+\frac{1}{2\pi}\int_{\omega_{2}}^{\infty}\frac{J(\omega)}{\omega-\omega_{1}}\mathrm{d}\omega,\\
\Omega_{2} & \equiv & \omega_{2}-\frac{1}{2\pi}\int_{-\infty}^{\omega_{1}}\frac{J(\omega)}{\omega_{2}-\omega}\mathrm{d}\omega+\frac{1}{2\pi}\int_{\omega_{2}}^{\infty}\frac{J(\omega)}{\omega-\omega_{2}}\mathrm{d}\omega,
\end{eqnarray*}
there will be a non-vanishing solution for $u(t)$ with oscillating
frequency $\omega_{0}$ in $(\omega_{1},\omega_{2})$. Especially
when the spectral density is discontinuous at $\omega_{1}$ and $\omega_{2}$,
e.g.,
\[
J(\omega)=\theta(\omega_{1}-\omega)\eta_{1}\mathrm{e}^{\gamma_{1}\omega}+\theta(\omega-\omega_{2})\eta_{2}\mathrm{e}^{\gamma_{2}\omega},
\]
$\Omega_{1}=-\infty$ and $\Omega_{2}=+\infty$, which implies that
the criterion (\ref{eq:37}) always holds. Thus, no matter how weak
the coupling strength is, such spectrum is always associated with
a non-vanishing $u(t)$ with oscillating frequency $\omega_{0}\in[\omega_{1},\omega_{2}]$.

\section{Influence of the environment}

The influence of the environment on the mean occupation number of
the system, 
\begin{equation}
\sum_{l}\left|u_{l}(t)\right|^{2}\left\langle \mathrm{b}_{l}^{\dagger}\mathrm{b}_{l}\right\rangle =\int p(\omega)f_{\beta}(\omega)\mathrm{d}\omega,
\end{equation}
can be viewed as the information {}``written'' into the system by
the environment. Here, 
\begin{equation}
p(\omega)=\sum_{l}\left|u_{l}(t)\right|^{2}\delta(\omega-\omega_{l}),
\end{equation}
is the distribution function (which usually are not normalized) and
$f_{\beta}(\omega)=1/\left[\exp(\beta\omega)-1\right]$ is the average
occupation number of the environment mode with frequency $\omega$
at temperature $T=1/(k_{B}\beta)$. 

To calculate $p(\omega)$, we first analyze the dynamic behavior of
$u_{l}(t)$. Following from Eq.~(\ref{eq:7}) and Eq.~(\ref{eq:9}),
we find that $u_{l}(t)$ obeys the differential-integral equation,
\begin{equation}
\frac{\mathrm{d}u_{l}(t)}{\mathrm{d}t}+\mathrm{i}\Omega u_{l}(t)+\int_{0}^{t}G(t-\tau)u_{l}(\tau)\mathrm{d}\tau=-\mathrm{i}\eta_{l}\mathrm{e}^{-\mathrm{i}\omega_{l}t},
\end{equation}
with initial condition $u_{l}(0)=0$. Comparing this equation with
Eq.~(\ref{du}), we express $u_{l}(t)$ in terms of $u(t)$ as\cite{YangLP2012}
\begin{equation}
u_{l}(t)=-\mathrm{i}\eta_{l}\int_{0}^{t}u(t-\tau)\mathrm{e}^{-\mathrm{i}\omega_{l}\tau}\mathrm{d}\tau.\label{eq:51}
\end{equation}
Thus $u_{l}(t)$ is determined by $u(t)$ over the time domain $[0,t]$.
As shown in the former sections, $u(t)$ decays exponentially in short
time and relaxes to an asymptotic form $A\exp(-\mathrm{i}\omega_{0}t)$
at lone time. For simplicity, we focus on the single-mode solution
case here and assume $u(t)$ to be the form,
\begin{equation}
u(t)\simeq\begin{cases}
\mathrm{e}^{-\mathrm{i}\Omega't-\gamma t}, & \ t<t_{1},\\
A\mathrm{e}^{-\mathrm{i}\omega_{0}t}, & \ t\geqslant t_{1}.
\end{cases}\label{eq:52}
\end{equation}
Though it seems a rough approximation, it seizes the essence. How
to choose a suitable $t_{1}$ will be pointed out later. Then we obtain
$u_{l}(t)$ as
\begin{eqnarray}
u_{l}(t) & = & -\mathrm{i}\eta_{l}\mathrm{e}^{-\mathrm{i}\omega_{l}t}\frac{\exp[-\mathrm{i}(\Omega'-\omega_{l})t_{1}-\gamma t_{1}]-1}{-\mathrm{i}(\Omega'-\omega_{l})-\gamma}-\mathrm{i}\eta_{l}A\mathrm{e}^{-\mathrm{i}\omega_{l}t}\frac{\exp[-\mathrm{i}(\omega_{0}-\omega_{l})t]-\exp[-\mathrm{i}(\omega_{0}-\omega_{l})t_{1}]}{-\mathrm{i}(\omega_{0}-\omega_{l})}.\label{eq:53}
\end{eqnarray}

Now, we consider two special cases: (1) $A=0$ for small $\eta$,
and (2) $A\not=0$ for large $\eta$. In the first case, the second
term in the right-hand side of Eq.~(\ref{eq:53}) vanishes, which
is equivalent to choose $t_{1}=\infty$ in Eq.~(\ref{eq:52}). Then,
the distribution function reads
\begin{eqnarray}
p(\omega) & = & \frac{1}{2\pi}\cdot\frac{J(\omega)}{(\omega-\Omega')^{2}+\gamma^{2}}.
\end{eqnarray}
The two parameters $\Omega'$ and $\gamma$ can be estimated by the
Wigner-Weisskopf approximation,
\begin{eqnarray}
\Omega' & = & \Omega-\frac{1}{2\pi}\int\mathrm{P}\frac{J(\omega')\mathrm{d}\omega'}{\omega'-\Omega},\\
\gamma & = & J(\Omega)/2.
\end{eqnarray}
It is discovered that $p(\omega)$ is a sharp distribution centered
at $\Omega'$ with width $\gamma$,
\begin{equation}
p(\omega)\simeq\frac{1}{2\pi}\cdot\frac{2\gamma}{(\omega-\Omega')^{2}+\gamma^{2}}.
\end{equation}
Since $A=0$, the first part of the system's mean occupation number
vanishes, thus the mean occupation number of the system reads
\begin{equation}
n(\gamma,\Omega')=\int\frac{1}{2\pi}\cdot\frac{2\gamma}{(\omega-\Omega')^{2}+\gamma^{2}}f_{\beta}(\omega)\mathrm{d}\omega.
\end{equation}
In the weak coupling strength limit $\eta\rightarrow0$ ($\gamma\rightarrow0$),
$p(\omega)\rightarrow\delta(\omega-\Omega')$, which leads to
\begin{equation}
n(\gamma\rightarrow0,\Omega')=\int\delta(\omega-\Omega')f_{\beta}(\omega)\mathrm{d}\omega=f_{\beta}(\Omega').
\end{equation}
Thus, the system with the renormalized frequency $\Omega'$ is thermalized
by the bath. If the coupling strength is small but finite, $p(\omega)$
becomes a Lorentzian-type distribution with broadened width and translated
center. The center translation effect is characterized by 
\begin{eqnarray}
\Delta n & =n(\gamma\rightarrow0,\Omega')-n(\gamma\rightarrow0,\Omega)\simeq & \frac{1}{2\pi}f_{\beta}^{\prime}(\Omega)\int\mathrm{P}\frac{J(\omega)\mathrm{d}\omega}{\Omega-\omega},
\end{eqnarray}
which is proportional to the coupling strength. The broadening width
effect is denoted by
\begin{eqnarray}
\Delta n & =n(\gamma,\Omega)-n(\gamma\rightarrow0,\Omega)\simeq & \frac{1}{2}f_{\beta}^{\prime\prime}(\Omega)\left\langle (\Delta\omega)^{2}\right\rangle ,
\end{eqnarray}
which is proportional to $\left\langle (\Delta\omega)^{2}\right\rangle \varpropto\gamma^{2}$. 

In the second case, $\gamma$ is very large ($\gamma\propto\eta$),
then $u_{l}(t)$ is dominated by the second term in the right-hand
side of Eq.~(\ref{eq:53}):
\begin{equation}
\left|u_{l}(t)\right|^{2}=A^{2}\left|\eta_{l}\right|^{2}\frac{2-2\cos(\omega_{0}-\omega_{l})(t-t_{1})}{(\omega_{0}-\omega_{l})^{2}}.
\end{equation}
Since $p(\omega)$ appears in the integral over $\omega$, the oscillation
term $\cos(\omega_{0}-\omega_{l})(t-t_{1})$ could be omitted. Then,
$p(\omega)$ is approximated as 
\begin{equation}
p(\omega)=\frac{A^{2}}{2\pi}\cdot\frac{2J(\omega)}{(\omega-\omega_{0})^{2}}.
\end{equation}